\documentclass[doublecol]{epl2} 
\usepackage{braket}
\usepackage{hyperref}
\usepackage{float}

\def\eps{\varepsilon}

\begin{document}

\title{Thermodynamic theory of voting and EU elections}

\author{Klaus M. Frahm \inst{1} \and  Dima L.Shepelyansky\inst{1}} 
\shortauthor{K. M. Frahm, D. L. Shepelyansky}
\institute{
\inst{1}Univ Toulouse, CNRS, Laboratoire de Physique Th\'eorique,
  Toulouse, France
}


\abstract{We introduce a thermodynamic theory of voting
  and show that it provides a good description of distribution
  of party votes in EU elections. The theory traces parallels between
  system energies of coupled nonlinear oscillators 
  and party vote fractions. Such a classical system evolution 
  is characterized by the conservation of total energy and
  probability norm that leads to the Rayleigh-Jeans  (RJ) thermalization
  and condensation at low energy states. A similar thermalization
  also describes the wealth inequality in society.
  This feature belongs to the phenomena of constraint driven condensation
  known in statistical mechanics.
  We show that the RJ theory well depicts
  the Lorenz and Pareto curves obtained from the EU vote results.
  The theory also recovers the dispersion of votes
  between candidates of first round presidential elections in France.
}



\maketitle

\section{Introduction}

In the European Union (EU) 
elections to the European Parliament take place every five years
by universal adult suffrage
with more than 400 million people eligible to vote \cite{euwiki,eueu}.
The campaigns of political parties
still take place through national elections
even if political parties have the right to campaign EU-wide.
The country election system
is a form of proportional representation of parties.
The electoral threshold, if any, may not exceed 5\%.
There is only one election round at which
a rather high number of parties, typically between 20 and 40, participate.

In this work, we use the EU public election results \cite{euwiki,eueu} and 
present their statistical distribution of party votes
for the time period from 1994 to 2024 
(for France and Germany) and for top 10 EU countries by
population for 2024 
(DE, FR, IT, ES, PL, RO, NL, BE, SE, CZ; here
countries are given by their alpha-2 codes ISO 3166-1).
We also consider the French 1st round presidential elections since 1965 
\cite{frpres}. 
To characterize the properties of this distribution,
we use a description based on the
Lorenz curve which gives the dependence of cumulated 
normalized wealth (vote fractions here) 
$0\leq w \leq 1$ on the cumulated normalized fraction of
population or households (political parties or electoral candidates here) 
$0 \leq h \leq 1$ \cite{lorenz}.
The case of perfect equipartition (identical vote fractions for all parties) 
corresponds to 
the diagonal $w=h$ and the doubled area between diagonal
and the Lorenz curve $w(h)$ gives the Gini coefficient
$0 \leq G \leq 1$  \cite{gini,boston} which 
is broadly used in the studies of wealth inequality 
in world  countries \cite{piketty1,piketty2,boston}.

The striking feature of wealth inequality is that for the whole
world 50\% of the population owns only 2\% of total wealth,
while 10\% (1\%) of the riches population owns 75\% (38\%) of the 
total wealth \cite{piketty2}.
In \cite{wth1,wth2} it was shown that this inequality can be naturally
explained by the  Rayleigh-Jeans (RJ) thermalization 
(see e.g. \cite{landau,zakharovbook}) that leads 
to RJ condensation (RJC) and the formation
of a huge poverty phase of low wealth and a tiny oligarchic phase
that captures the main part of total society wealth.
It is also argued that the RJC phenomenon
describes the energy and carbon emission
distribution between the world countries \cite{enth}.

In physical systems RJC is also observed in multimode optical
fibers \cite{wabnitz,picozzi1,chrisrep,picozzi2,ourfiber}.
This is a specific case of
a more generic phenomenon known in statistical mechanics as
constraint-driven condensation \cite{trizac,satya,marsili}
that is universal and
exists for such systems as
coalescence in granular media, jamming in traffic, 
gelation in networks \cite{satya}
and financial data analysis \cite{marsili}.
RJC also exists in systems of dynamical chaos
where  moderate nonlinear couplings between
oscillators lead to RJ thermalization \cite{wth1,wth2,ourfiber,stratif}.

In previous works RJ thermalization and condensation
were considered for some physical objects
like light intensity in fibers, wealth, energy
or carbon emission \cite{wth1,wth2,enth}.
Here we develop the Thermodynamic Theory of Voting (TTV)
according to which opinion formation
and Vote Preferences (VP) result from
certain complex interactions between agents/electors in a society
that leads to a thermal distribution of votes
between parties participating in elections.
Of course, it is assumed that the number of parties
is significantly higher than two.
We show that the TTV approach gives a good
description for the party/competitors distribution
of votes in EU elections and also in
the first round presidential elections in France. 

We point out that there is a developed mathematical theory
of voting with many nontrivial results
(see e.g. \cite{vote1,vote2,vote3,vote4}).
But there the electors are usually considered as
independent noninteracting agents.
In contrast to this, in the TTV approach it is assumed that there are certain
couplings and nonlinear interactions between
electors operating in the framework of classical mechanics. 
This leads to thermalization of voting
with the emergence of RJ distribution of votes
between parties. In a certain sense this process
is similar to the thermalization of interacting atoms
in a three-dimensional box: almost any interactions lead
to a thermal steady-state with average energy
$3T/2$ given by temperature $T$ \cite{landau}
with the Boltzmann constant taken here as unity
(but no themalization happens without nonlinear interactions).
We show that the TTV concept provides a detailed statistical
description of real results of EU and other elections.

We also note that there are various physical models
of opinion formation and voting of interacting agents
(see e.g. \cite{sznajd,redner,galam,bukina} and Refs. therein)
but they do not yet describe the results of real elections.

\section{RJC and Lorenz, Pareto curves construction}

Yakovenko et al. \cite{yakovenko1,yakovenko2} proposed to use 
the statistical Boltzmann-Gibbs (BG) approach to describe the 
distribution of money, wealth and income in terms of an exponential 
function depending on the temperature parameter $T$. 
However, it happens that in this case  the Lorenz curve
does not depend on $T$ and has always the same Gini coefficient $G=0.5$
\cite{yakovenko2,wth2}. Therefore, in \cite{yakovenko2}, an extension 
was proposed where the tail of the BG distribution is 
replaced by a power law tail, according to the Pareto distribution. 
This procedure appears to be somewhat artificial and does not seem to 
be natural. 

A different view point is given by
the Wealth Thermalization Hypothesis (WTH), introduced in \cite{wth1,wth2},
which is based on the idea of the conservation of two quantities as it 
was argued in \cite{boghosian1,boghosian2} in the context of wealth 
distribution. 
This approach assumes that the wealth in a country is analogous to 
the global energy 
of a certain (chaotic) classical oscillator system with specific weak 
nonlinear interactions between the oscillators such that there 
are indeed two integrals of motion which are the global energy $E$ 
and the global norm, given as the sum of all squared oscillator 
amplitudes. Such a system thermalizes under certain conditions \cite{wth1} 
and in this case for each oscillator at energy (frequency)  $E_m$, the 
statistical average of its norm probabilities (squared amplitudes)
is given by the 
RJ expression \cite{landau,zakharovbook}:
\begin{equation}
\rho_m = \frac{T}{E_m-\mu} \; ({\rm RJ})\ .
\label{eqrj}
\end{equation}
Historically, 
this expression describes the thermalization of classical fields. 
In the context of the WTH, $\rho_m$ represent averaged stationary 
probabilities at certain wealth states 
 $0 \leq m < N$ with wealth/energies $w_m=E_m$. 
The two parameters $T$ and $\mu$ are the system temperature $T$ and 
the chemical potential $\mu(T)$ which are 
determined by the two implicit equations $E= \sum_m E_m \rho_m$ 
and $\eta= \sum_m \rho_m =1$ (at a given value of the global 
energy $E$ and for a given oscillator spectrum $E_m$; see \cite{wth1} 
for details). 

In this work, we extend the ideas of the WTH to elections (TTV approach) 
where the vote fractions $V_m$ correspond to wealth or energy 
$w_m = E_m = V_m$ (in the following, we usually use equivalently 
these three quantities) and political parties or election candidates 
corresponding to (vote fractions) of population (or occupation probabilities 
for certain oscillator modes). 

The important feature of the RJ distribution (\ref{eqrj})
is the possible condensation of a macroscopic part of the 
total probability $\eta$ 
at the lowest energy state with $E_0$ (or few lowest states $E_m$) 
provided the global energy or temperature is sufficiently low, 
$E\sim E_1-E_0$ or $\eps\ll 1$. Here 
$\eps=E/B$ is the rescaled energy and $B=E_{N-1}$ is the energy bandwidth 
(in this work, we assume $0=E_0<E_1<\ldots<E_{N-1}=B$ and we usually choose 
models with $B\approx 1$). 

\begin{figure}[h]
\begin{center}
\includegraphics*[width=8.5cm]{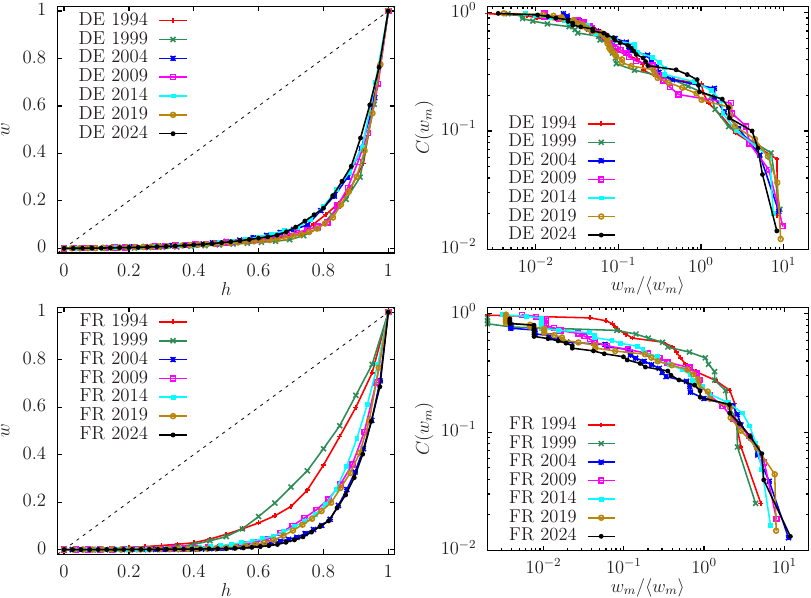}
\end{center}
\vglue -0.3cm
\caption{(Color on-line) 
{\em Left:} Lorenz curves of EU elections of Germany (DE; top) and France 
(FR, bottom) since 1994. The $x$-axis corresponds to the 
cumulated fraction of households/political parties ($h$) and the $y$-axis to
the cumulated fraction of wealth/obtained votes ($w$). 
The dashed black line 
corresponds to the line of perfect equipartition $w=h$. 
The Gini coefficients $G$ and number of political parties $N_p$ 
for all cases can be found in Table~\ref{tab1}. 
{\em Right:} Pareto curves $C(w_m)$ for the same data where 
$C(w_m)$ represents the fraction of parties with a vote fraction 
larger than $w_m$ (analogous to wealth). 
The $x$-axis corresponds to rescaled 
values $w_m/\langle w_m\rangle$ where $\langle w_m\rangle = 1/N_p$ is the 
average vote fraction (i.e. assuming $w_m$ units such that $\sum_m w_m=1$). 
}
\label{fig1}
\end{figure}
For a given set of wealth values/vote fractions $w_m=E_m=V_m$, 
we construct the Lorenz curve as the set of points $(h(m),w(m))$ 
($m=0,1,\ldots,N$) where $h(m) =\sum^{m-1}_{k=0} \rho_k $
are the cumulated normalized fractions of households/parties
and $w(m) = \sum^{m-1}_{k=0} w_k \rho_k/w_s$ 
are the cumulated wealth/vote fractions. 
Here, $w_s = \sum_m w_m\rho_m$ is the average wealth/vote fraction such 
that $h(0)=w(0)=0$ and $h(N)=w(N)=1$. 

For the case of real election data, we use 
$\rho_m =1/N_p$ and vote fractions $w_m=V_m=E_m$ 
taken from the databases \cite{euwiki,eueu,frpres}.
Here $N_p$ is the number of political parties/candidates having 
participated at a certain election of a given year. For EU elections, 
since 1994, we 
have mostly $20\leq N_p\leq 41$ (with a few cases of $N_p\sim 13-15$) 
and for the first round FR presidential election since 1965 we have 
$6\le N_q\le 16$. (see Table~\ref{tab1}). 

For the case of a thermalized RJ model, using a certain simple model 
spectrum $E_m$ with $0=E_0<E_1<\ldots<E_{N-1}=B$ with some large 
value $N\ge 10^4$ (or 
even $N\to\infty$ \cite{wth1,wth2}), we compute the Lorenz curve 
in the same way using for $\rho_m$ the RJ expression 
(\ref{eqrj}) and for $w_s=E=\eps B$ the global energy which is a given 
parameter. In this case, it is assumed that there are $N$ thermalized agents 
with energy levels $E_k$ for their political power in society and that 
each party is represented by a certain fraction of them. 

We mention that a possible condensation (for very low values of $\eps\ll 1$) 
where a significant fraction 
of probability is concentrated on a few modes (or only one mode at $m=0$) 
corresponds to $w(h)\approx 0$ for a rather large interval of $h\in[0,h_c]$.
This naturally explains the appearance of a strong
wealth inequality when a huge part of population is poor 
and a tiny oligarchic population 
fraction captures a big part of total wealth 
as it is described in \cite{wth1,wth2,enth}. 
We show that a somewhat similar situation 
may take place for the party vote distribution 
in certain EU elections. 

We note that the construction procedure of Lorenz curves is invariant with 
respect to global rescaling of $E_m\to\alpha E_m$ and $E\to \alpha E$. 
Therefore, we can choose model spectra with $B=1$ such that $\eps=E$. 
In \cite{wth1,wth2}, two specific spectral models were considered. 
The first one is based on the simple assumption that 
the density of energy states
$\nu(k) =d k/d E_k$ is constant such that
$0 \leq E_k=k/N < B=1$ for $k=0,1,2,...,N-1$. This case 
is called the RJ standard (RJS) model \cite{wth1,wth2,ourfiber}.
The second case is the RJ extended (RJE) model 
with $E_k=(e^{ak/N}-1)/(e^a-1)$ where $a$ is an additional real valued 
model parameter (the limit $a\to 0$ reproduces the RJS model). 
For $a>0$, the density of states decreases at high energies
with $\nu(E_k)=dk/dE_k = N(e^a-1)/[a(1+(e^a-1)E_k]$. In both cases, the global 
energy scale is chosen such that $B=E_{N-1}\approx 1$. 

For many typical cases studied in \cite{wth1,wth2}, e.g. Lorenz curves 
of household wealth in the World or countries, GPD of countries, 
stock market capitalization of companies etc., 
the RJE model describes very well real curves with typical optimal values 
of  $a\sim 4$. In these cases, the 
simple RJS model (i.e. $a=0$) gives only an approximate 
agreement woth data while the RJE curve typically 
matches the real data very well. 
However, in this work, for the cases of election Lorenz curves,
we usually have  small $a$ values and the 
resulting Lorenz curves for the RJE model are very close to the RJS curves. 
We point that as in \cite{wth1,wth2}, when comparing real data with the 
RJS or RJE model, the value of $\eps$ is determined such that the Gini 
coefficient of the RJ model coincides with its value from the real data.

\begin{figure}[h]
\begin{center}
\includegraphics*[width=8.5cm]{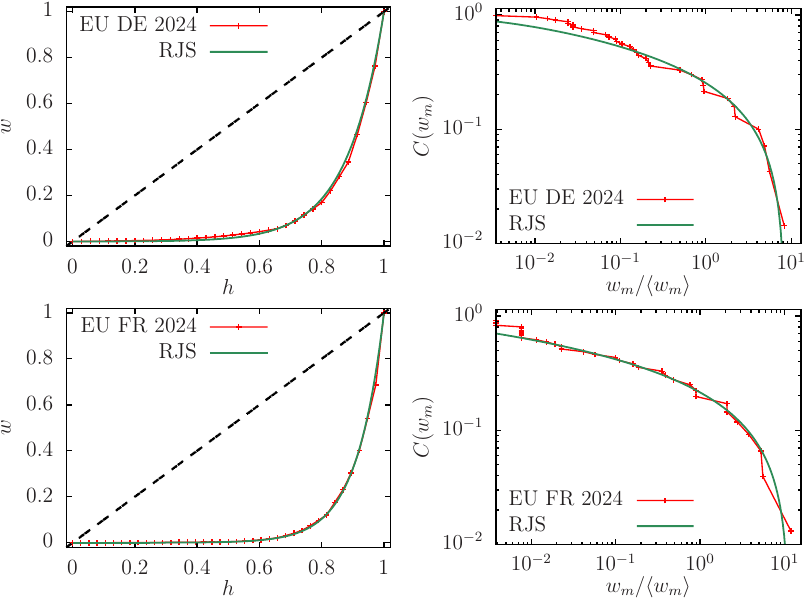}
\end{center}
\vglue -0.3cm
\caption{(Color on-line) 
As Fig.~\ref{fig1} for the EU 2024 elections of DE (top) and 
FR (bottom). In addition, to the real data (red points) also the 
(green) theoretical RJS Lorenz curves (left panels)
are shown (see Table~\ref{tab1} for specific values of $G$, $\eps$ 
and $N_p$). For the Pareto curves (right panels) the 
average $\langle w_m\rangle$ 
is either $1/N_p$ (real data) or $\eps$ (RJS curves). 
See also SupMat Figs. S1 and S2 for similar figures
for the DE/FR EU elections 
of 1994 (Fig.~S1) and 2014 (Fig.~S2)
(these Figs. also show curves for RJE model
being  very close to the RJS ones). 
}  
\label{fig2}
\end{figure}

In addition to the Lorenz curve, we also study the 
cumulative distribution function (CDF) $C(w_m)$
that presents the fraction of households (parties) having 
a wealth/vote larger than $w_m=V_m$. For convenience, we call this 
function Pareto distribution (Pareto curve), even for more general 
cases where it does not obey an expected power law. 
The Pareto curve highlights in a better way
the properties of high revenues (see e.g. \cite{yakovenko2}).
For both cases of real data and RJ models, the Pareto curve 
is easily obtained by drawing $C(w_m) = 1- h(w_m)$ 
as a function of the individual household wealth less than $w_m$.
Here $h(w_m)$ corresponds to
the fraction of (poorest) households having each 
an individual household wealth less than $w_m$. 
However, in contrast to the Lorenz curve, the Pareto curve still depends 
on a scale parameter for the units of $w_m$. In this work, 
we always show $C(w_m)$ as a function of the rescaled quantity 
$w_m/\langle w_m\rangle$ where $\langle w_m\rangle$ is the average 
``wealth'' which is either directly known for real data or given 
by $\eps$ (for both RJ models with $B=1$). For the case of elections, 
we simply have $\langle w_m\rangle=1/N_p$ if $w_m$ represents vote 
fractions normalized by $\sum_m w_m=1$. 

We mention, that the practical computation of Lorenz and Pareto curves 
for both RJ models (at given values of $\eps$ and $a$), can be done very 
efficiently for quite large spectra $N\ge 10^4$ using the above construction 
procedure and numerically obtained values for $\mu$ and $T$ as a function 
of $\eps$. Furthermore, for the continuous limit 
$N\to\infty$ also explicit analytic formula for both $w(h)$ and $C(w_m)$ 
are available (see \cite{wth2} for details). The RJ curves shown in this 
work were all computed for this continuous limit but curves for 
finite $N=10^4$ are identical on graphical precision (with relative 
deviations $\sim 10^{-4}$). 

\section{Data}

In this work, we consider all 7 EU elections of DE and FR 
between 1994 and 2024, the latest 2024 EU elections of other 8 countries, 
IT, ES, PL, RO, NL, BE, SE, CZ (decreasing order of population) 
and also all 11 French 1st round presidential elections since 1965 using 
data of \cite{euwiki,eueu,frpres}. 

The main parameters of these elections, being the Gini coefficient 
$G$, obtained RJS value of $\eps$ from the given $G$ value of each election, 
and the number $N_p$ of political 
parties or French presidential first round candidates having participated 
are given in Table~\ref{tab1}.

\setlength{\arrayrulewidth}{.1em} 
\begin{table}[h]
\caption{\label{tab1}Summary of parameters of certain EU and 
(first round) FR presidential elections (PR FR). The columns correspond to the 
type of election, year of election, Gini coefficient $G$, 
RJS rescaled energy $\eps$, obtained by matching $G$ between 
real data and the theoretical RJS Lorenz curve, and the number $N_p$ 
of political parties 
or (first round) presidential candidates having participated at the 
election. 
}
\begin{center}
\begin{tabular}{lcccr}
\hline
type & year & $G$ & $\eps$ & $N_p$\\
\hline
EU DE & 1994 & $0.802$ & $0.099$ & $26$\\
EU DE & 1999 & $0.820$ & $0.090$ & $23$\\
EU DE & 2004 & $0.772$ & $0.115$ & $24$\\
EU DE & 2009 & $0.800$ & $0.100$ & $32$\\
EU DE & 2014 & $0.766$ & $0.118$ & $25$\\
EU DE & 2019 & $0.810$ & $0.095$ & $41$\\
EU DE & 2024 & $0.763$ & $0.119$ & $35$\\
\hline
EU FR & 1994 & $0.635$ & $0.194$ & $20$\\
EU FR & 1999 & $0.597$ & $0.219$ & $20$\\
EU FR & 2004 & $0.816$ & $0.092$ & $39$\\
EU FR & 2009 & $0.765$ & $0.118$ & $27$\\
EU FR & 2014 & $0.752$ & $0.125$ & $31$\\
EU FR & 2019 & $0.783$ & $0.109$ & $34$\\
EU FR & 2024 & $0.824$ & $0.088$ & $38$\\
\hline
EU IT & 2024 & $0.625$ & $0.200$ & $15$\\
EU ES & 2024 & $0.848$ & $0.076$ & $34$\\
EU PL & 2024 & $0.654$ & $0.181$ & $20$\\
EU RO & 2024 & $0.658$ & $0.179$ & $13$\\
EU NL & 2024 & $0.574$ & $0.236$ & $20$\\
EU BE & 2024 & $0.551$ & $0.255$ & $22$\\
EU SE & 2024 & $0.696$ & $0.156$ & $20$\\
EU CZ & 2024 & $0.769$ & $0.116$ & $30$\\
\hline
PR FR & 1965 & $0.530$ & $0.272$ & $6$\\
PR FR & 1969 & $0.548$ & $0.257$ & $7$\\
PR FR & 1974 & $0.742$ & $0.130$ & $12$\\
PR FR & 1981 & $0.539$ & $0.264$ & $10$\\
PR FR & 1988 & $0.507$ & $0.292$ & $9$\\
PR FR & 1995 & $0.404$ & $0.403$ & $9$\\
PR FR & 2002 & $0.456$ & $0.343$ & $16$\\
PR FR & 2007 & $0.613$ & $0.208$ & $12$\\
PR FR & 2012 & $0.560$ & $0.247$ & $10$\\
PR FR & 2017 & $0.548$ & $0.257$ & $11$\\
PR FR & 2022 & $0.573$ & $0.237$ & $12$\\
\hline
\end{tabular}
\end{center}
\end{table}

\section{Results}

In Fig.~\ref{fig1}, we show Lorenz and Pareto curves
for Germany (top row) and France (bottom row)
for 7 EU elections from years 1994 to 2024.

For Germany these distributions remain stable 
and there are only modest variations for the full period. 
The number of competing parties is between 23 (1999) and 41 (2019). 

\begin{figure}[h]
\begin{center}
\includegraphics*[width=8.5cm]{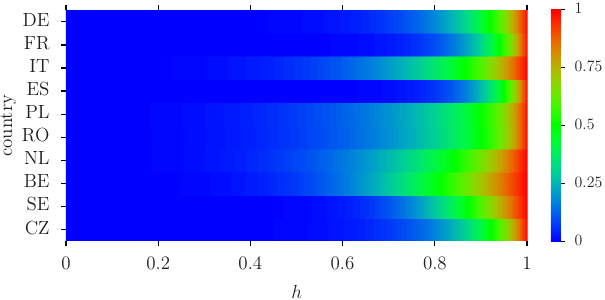}
\end{center}
\vglue -0.3cm
\caption{(Color on-line) 
Density color plot of Lorenz curves for 10 countries (2024 EU elections) 
obtained from the RJS model for the 6 countries 
DE, FR, IT, ES, NL, CZ and from the optimal RJE model for the 
4 countries PL, RO, BE, SE (cases with  a visible difference 
between RJS and optimal RJE curve). 
The $x$-axis corresponds to the 
cumulated fraction of households/political parties ($h$) and the $y$-axis to
the countries (with order of increasing population from bottom to top). 
The color value according to the colorbar shows  the  
cumulated fraction of wealth/obtained votes ($w$). 
See also SupMat Figs. S3-S6 for figures in the style of Fig.~\ref{fig2} 
with Lorenz and Pareto curves for the 2024 EU elections of 8 EU countries 
(CZ to IT). )
}  
\label{fig3}
\end{figure}

For France the curves for 1994 and 1999 
have visible deviations and smaller Gini coefficients than 
the other years which may be related to the reduced number of 
political parties $N_p=20$ for these two years while for the 
other years $N_p\ge 27$ with maximal value $N_p=39$ for 2004. 

It is important to note that the Pareto curves in Fig.~\ref{fig1} 
clearly do not show an expected algebraic decay over a significant interval 
of $w_m$ (see e.g. \cite{yakovenko2} and Refs. therein). 

In global, these Lorenz and Pareto curves remain stable on a scale of 
the last 30 years for Germany and 20 years for France that
indicates a certain stable mechanism behind. Below, we present arguments
that this mechanism is related to RJ thermalization and condensation. 

For example for the 2024 EU elections of DE and 
FR, Fig.~\ref{fig2} shows that the real data Lorenz and Pareto curves 
agree very well with the corresponding theoretical curves obtained from 
the RJS model using the above construction procedure with the RJ 
expression (\ref{eqrj}) for $\rho_m$ and a uniform density of states for the 
spectrum $E_m$. 
This is also confirmed, by the similar SupMat Figs.~S1 and S2 for 
1994 and 2014 EU elections of DE and FR;
also for these years RJS and RJE curves are very close.

\begin{figure}[h]
\begin{center}
\includegraphics*[width=8.5cm]{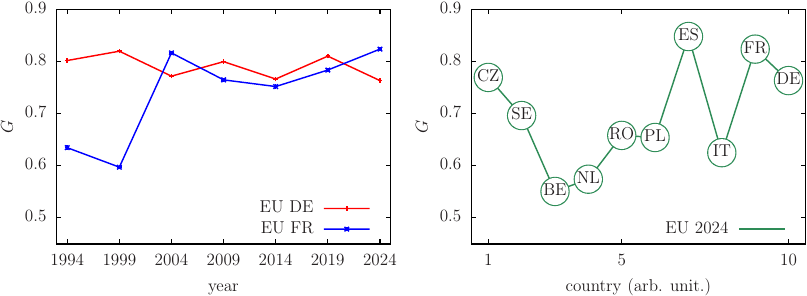}
\end{center}
\vglue -0.3cm
\caption{
(Color on-line) 
{\em Left:} Dependence of the Gini coefficient $G$ for the EU elections 
of DE (red) and FR (blue) on the election year (using the same data 
as in Fig.~\ref{fig1}). 
{\em Right:} Gini coefficient $G$ for the EU elections 
2024 for 10 European countries (green) 
ordered with increasing size of population (from left to right; using 
the same data as in ~Fig.~\ref{fig3}). 
}  
\label{fig4}
\end{figure}

In Fig.~\ref{fig3}, we show a collection of all Lorenz curves 
for the 2024 EU elections of the 10 most populated EU countries 
for either the RJS model (6 cases) or the RJE model with optimal 
value of $a$ (4 cases) as a density color plot. For all these cases, 
the shown RJ curves agree very well 
with the real data curves as can be seen in SupMat Figs.~S3-S6 
(for the 4 cases with RJE curves in Fig.~\ref{fig3} 
there is some visible difference between 
the RJE/real data curve and the corresponding RJS curve and here we have 
larger optimal values of $a$ with $|a|>2$). Here the number of political 
parties $N_p$ is rather small for RO (13) and IT (15), 
intermediate for PL, NL, BE and SE (20-22) and somewhat larger 
for CZ (30), ES (34), DE (35) and FR (38). 

Fig.~\ref{fig4}, shows the dependence of the Gini coefficient $G$ 
for the DE/FR EU elections 
on the election year between 1994 and 2024 (left panel) 
and for the 2024 EU elections of the 10 most populated EU countries on the 
country index (defined by order of increasing population). 

For FR 1994/1999 there are some deviations with smaller values of 
$G\approx 0.6$, also visible in Fig.~\ref{fig1} on a qualitative level,
and corresponding RJS values of $\eps\approx 0.2$. For the other years of 
FR and all years of DE, we have typically $G\approx 0.8$ and 
$\eps\approx 0.1$, 
indicating lower temperatures and the emergence of a (stronger) RJ 
condensation. 
In particular, from the RJS Lorenz curves of Fig.~\ref{fig2} 
we see for DE (FR) that 50\% of bottom parties gain approximately only 
1.4\% (0.3\%) of total votes while top 10\% of parties 
gain approximately 57\% (68\%) of total
votes. These values are  comparable with
those of wealth inequality for the whole world in 2021 
with $G=0.842$ \cite{piketty2,wth1}. However, the World 2021 Lorenz 
curve analyzed in \cite{wth2} requires a large RJE parameter $a=4.74$ and 
there the RJS curve does not match very well the real data,
In contrast 
for the FR/DE 2024 EU election Lorenz curves are well described by
the RJS model.. 

For the other 8 EU elections of 2024 the Gini coefficients 
vary between $G\approx 0.55$ (BE with $\eps\approx 0.26$) 
and $G\approx 0.85$ (ES with $\eps\approx 0.076$). Globally, $G$ decreases 
with increasing $\eps$ (increasing temperature). The analytic theory of 
the RJS model for $N\to\infty$ \cite{wth1,wth2} gives for 
$\eps\le 0.2$ the quite good 
approximation $G\approx 1-2\eps$ (the complete data of SupMat Table S1 
is quite well fitted by $G\approx 1-2\eps+3.4\,\eps^3$ for the larger 
interval $\eps\in[0,0.4]$). 

\begin{figure}[h]
\begin{center}
\includegraphics*[width=8.5cm]{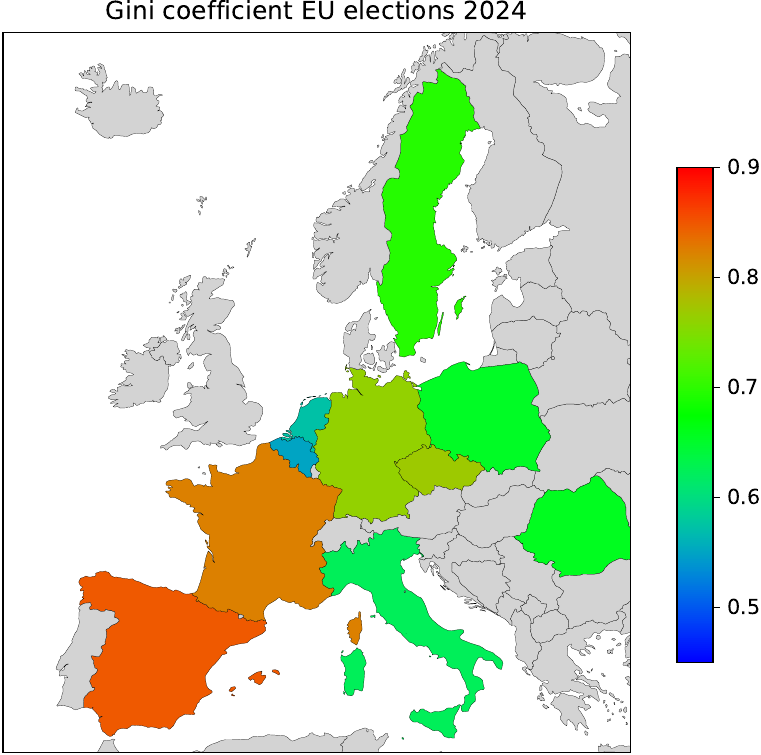}
\end{center}
\vglue -0.3cm
\caption{(Color on-line) 
Europe map color plot of the Gini coefficient of the EU elections of 2024 of  
10 countries using the data of the right panel of Fig.~\ref{fig4}.  
}  
\label{fig5}
\end{figure}

The Gini coefficients of the 10 most populated EU countries are 
also illustrated in the Europe map color plot of Fig.~\ref{fig5}. 
We note that the values of $G$ for the 10 
countries, visible in the right panel of Fig.~\ref{fig4} 
and in Fig.~\ref{fig5}, 
also agree on a qualitative level with the RJ Lorenz curves in the color 
plot of Fig.~\ref{fig3}, i.e. the length of the green/red zone 
in Fig.~\ref{fig3} increases with decreasing values of $G$. 

\begin{figure}[h]
\begin{center}
\includegraphics*[width=8.5cm]{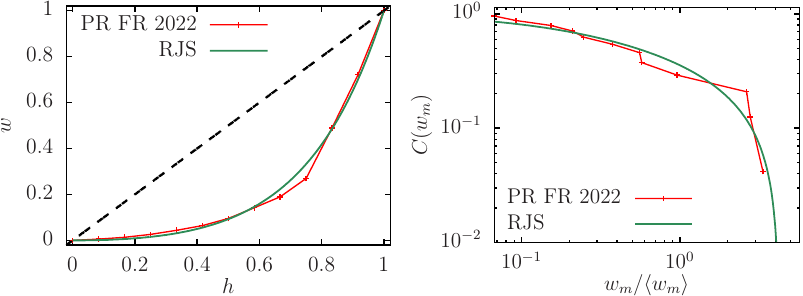}
\end{center}
\vglue -0.3cm
\caption{(Color on-line) 
Same style as Fig.~\ref{fig2} but for the first round FR presidential 
election of 2022. 
See also SupMat Figs. S7-S11 for similar figures for the 
first round FR presidential elections between 1965 and 2017 (10 elections) 
and SupMat Fig. S12 for the dependence of the Gini coefficient 
of all FR first round presidential since 1965 on the election year 
(11 elections). 
In all cases, the parameter values for $G$, $\eps$ and $N_p$ are 
available in Table~\ref{tab1}.
}  
\label{fig6}
\end{figure}

Finally, it is important to note
that the TTV approach works not only
for elections of parties but 
also for elections of individual candidates if their number is not too small. 
As examples, we consider the case of the first round French presidential 
elections using data from \cite{frpres}, where typically a significant 
number of candidates participate, 
even though this number, typically $9$-$12$ for most years and 
with extreme values between 6 (1965) and 16 (2002), is quite 
smaller than the number of parties of typical EU elections. 

However, Fig.~\ref{fig6} still shows a very good agreement between the 
real data and the theoretical RJS Lorenz and Pareto curves for the 2022 
election. SupMat Figs. S7-S11 provide rather similar results for the other 
10 elections from 1965 to 2017. The dependence of the Gini coefficients 
for the 11 first round presidential elections between 1965 and 2022 on 
the election year is shown in SupMat Fig.~S12. Here, typically $G$ is 
mostly in the interval $G\in [0.45,0.6]$ with exceptions for 
1995 ($G\approx 0.4$) and 1974 ($G\approx 0.74$). These values are 
globally quite 
smaller as for most EU elections, which can be explained by the reduced value 
of $N_p$ such that a strong inequality is less likely. The dependence of 
$G$ on the RJS rescaled energy $\eps$ (obtained by matching $G$ between the 
real data and the RJS Lorenz curves) is the same as already given above, but 
since most typical $G$ (or $\eps$) values are smaller (larger) in comparison 
to the EU elections, the more general 3rd order fit needs to be used 
for most data points. 


\section{Discussion}

In this work, we introduce the TTV approach which 
describes very well the statistical distribution
of votes for parties participating in
EU elections during the last 30 years and for candidates 
of the first round French presidential elections since 1965. 
The theory is based on RJ thermalization with possible condensation 
which has been observed in other physical systems such as in multimode 
optical fibers \cite{wabnitz,picozzi1,chrisrep,picozzi2,ourfiber}. 
The RJ condensation appears for a low total system energy 
when a macroscopic fraction of probability is concentrated 
in a vicinity of the ground state energy.
A similar phenomenon is also known as the constraint driven
condensation in statistical mechanics of various
classical systems \cite{trizac,satya,marsili}. 
In the context of wealth inequality in the world, 
this RJ condensation \cite{wth1,wth2} explains a huge poverty phase and
a tiny oligarchic phase observed for 
many world countries \cite{piketty1,piketty2}. More 
recently, it was shown that the RJ approach also describes well 
energy and carbon emission distributions
between countries \cite{enth}. In this work, we extended this approach
to the statistical distribution of vote fractions in elections 
confirming that it describes well 
EU and French first round presidential elections. 
In the context of elections a possible RJ condensation corresponds 
to the appearance of a large fraction of parties
that cumulate a very small percent of total votes while 
few top parties accumulate a big fraction of total votes. 

The observation that RJ thermalization, applies   
to many statistical steady-state distributions in
physical and social systems confirms its universal 
nature for these systems. 

The TTV approach can be also applied 
to other types of elections, 
e.g. for elections of various committees 
in companies, or local elections. 
Of course, the TTV description 
only provides a general physical constraint
for typical vote distributions but does not tell who will win an election.

{\bf Acknowledgments:} The authors acknowledge support from the grant
 ANR France project
NANOX $N^\circ$ ANR-17-EURE-0009 in the framework of 
the Programme Investissements d'Avenir (project MTDINA).
This work was granted access to the HPC resources of
CALMIP (Toulouse) under the allocation 2026-P0110. 

{\bf Data Availability Statement:} The research data associated with
this article are included within the article and the used raw data 
is available in \cite{euwiki,eueu,frpres}.

\newpage

\noindent{\bf
Supplementary Material for \\ 
\vskip 0.2cm
\noindent Thermodynamic theory of voting and EU elections}

\bigskip

\noindent by
K.~M.~Frahm and D.~L.~Shepelyansky\\
\noindent Laboratoire de Physique Th\'eorique, 
Universit\'e de Toulouse, CNRS, UPS, 31062 Toulouse, France
\bigskip

Here, we present additional figures and additional information. 
We remind that Table~1, gives for all elections discussed in this work 
a summary of the 3 parameters $G$ (Gini coefficient), 
$\eps$ (rescaled energy for the RJS Lorenz curve obtained by matching 
$G$ with the real data Lorenz curve) and the number $N_p$ 
of political parties (EU elections) or candidates for the (first round) 
FR presidential election. Since these values are available in 
Table~1, they are not mentioned in the figure captions of 
the figures of the main part or other figures given below. 

In certain cases, we provide for illustration or due to visible deviations 
between the RJE and RJS cases also RJE Lorenz and Pareto curves with 
values of the optimal parameter $a$ (which minimizes the geometrical Lorenz 
curve distance with respect to the real data) and the associated value 
of $\eps_{\rm RJE}$ (which may deviate from $\eps$ of the RJS model given 
in Table~1). 

The following figures concern (mostly) additional figures in the style of 
Fig.~2 (except Fig.~\ref{figS12}; see below). 

More specifically, 
Figs. \ref{figS1} and \ref{figS2} concern 
the EU elections of DE/FR of 1994 (Fig.~\ref{figS1}) 
and of 2014 (Fig.~\ref{figS2}) both also with RJE curves (blue) for 
illustration. 

Figs. \ref{figS3}-\ref{figS6} concern the EU elections of 2024 of 8 countries 
other than DE or FR (from the list of 10 countries used in the main part 
and in Figs. 3-5). Here, in four cases with visible deviations between 
the RJE and RJS models, also curves and parameter values for the RJE case 
are given (see also caption of Fig.~3 for the list of countries where RJE 
curves are shown). 

Figs.~\ref{figS7}-\ref{figS11} concern the 10 first round FR presidential 
elections of the years 1965 to 2017 (the case of 2022 is shown in Fig.~6 
of the main part).

Finally, Fig.~\ref{figS12} shows the dependence of the Gini coefficient 
on the election year for the 11 first round FR presidential 
elections of the years 1965 to 2022. 

\renewcommand{\thetable}{S\arabic{table}}
\setcounter{figure}{0} \renewcommand{\thefigure}{S\arabic{figure}}

\begin{figure}[h]
\begin{center}
\includegraphics[width=0.95\columnwidth]{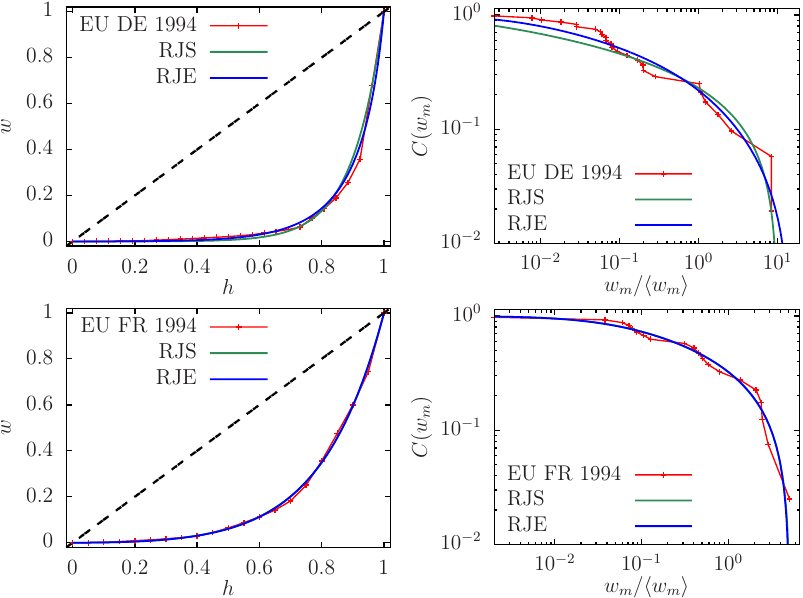}%
\end{center}
\caption{\label{figS1}
As Fig.~2 for the EU election of DE/FR of 1994. 
For illustration also the theoretical RJE curves are shown (blue). 
The optimal RJE parameters are $a=1.25,\ \eps_{\rm RJE}=0.069$ (EU DE 1994) 
and $a=-0.0701,\ \eps_{\rm RJE}=0.196$ (EU FR 1994).
}
\end{figure}

\begin{figure}[h]
\begin{center}
\includegraphics[width=0.95\columnwidth]{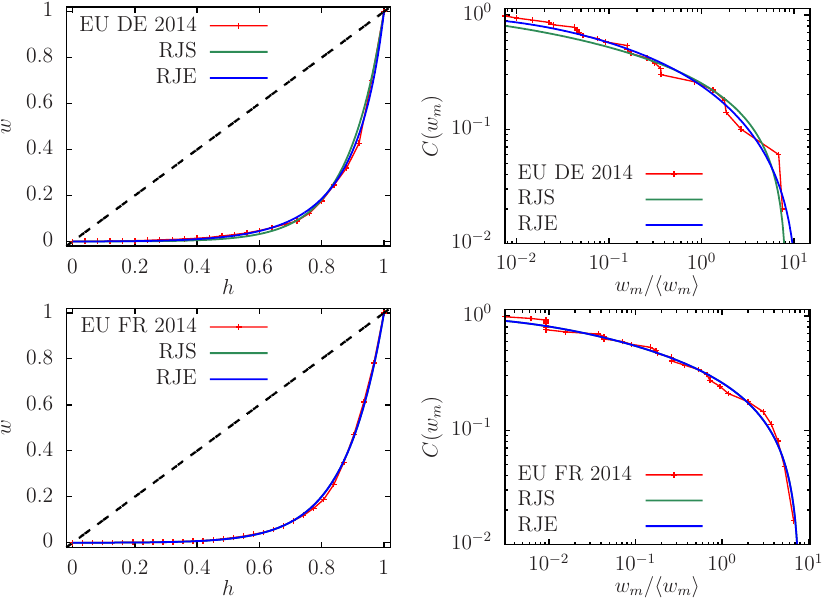}%
\end{center}
\caption{\label{figS2}
As Fig.~2 for the EU election of DE/FR of 2014. 
For illustration also the theoretical RJE curves are shown (blue). 
The optimal RJE parameters are $a=1.07,\ \eps_{\rm RJE}=0.0875$ (EU DE 2014) 
and $a=0.00573,\ \eps_{\rm RJE}=0.125$ (EU FR 2014).
}
\end{figure}

\begin{figure}[h]
\begin{center}
\includegraphics[width=0.95\columnwidth]{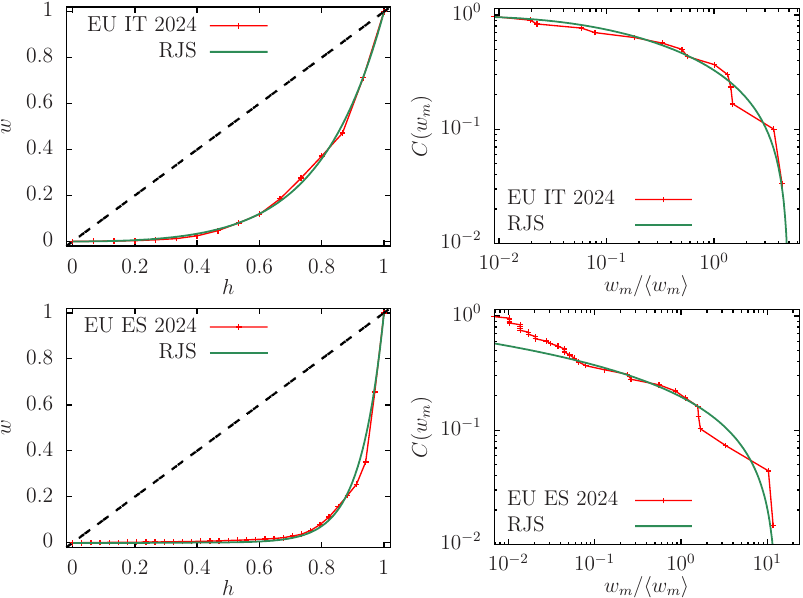}%
\end{center}
\caption{\label{figS3}
\label{fig_euvoteIT2024}
As Fig.~2 for the EU election of IT/ES of 2024. 
}
\end{figure}

\begin{figure}[h]
\begin{center}
\includegraphics[width=0.95\columnwidth]{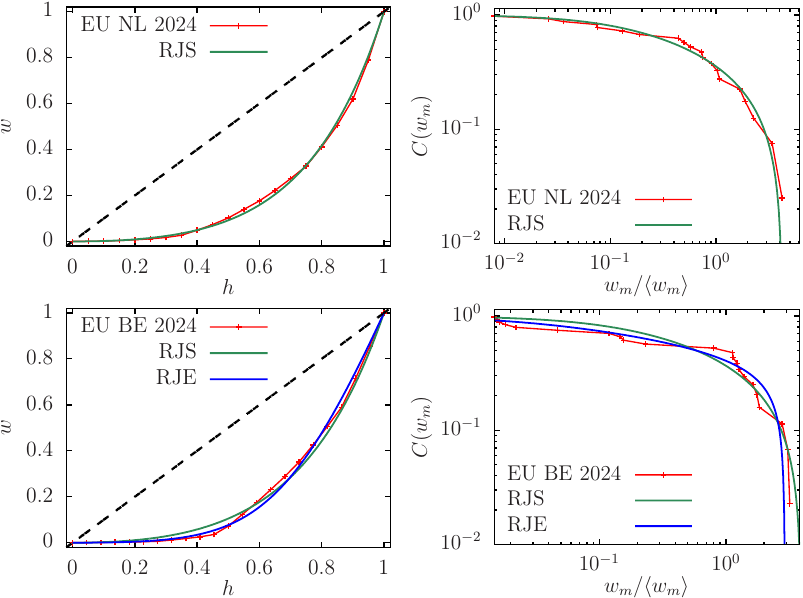}%
\end{center}
\caption{\label{figS4}
\label{fig_euvoteNL2024}
As Fig.~2 for the EU election of NL/BE of 2024. 
For BE also the theoretical RJE curves are shown (blue) and 
the optimal RJE parameters are $a=-2.35,\ \eps_{\rm RJE}=0.342$. 
}
\end{figure}

\begin{figure}[h]
\begin{center}
\includegraphics[width=0.95\columnwidth]{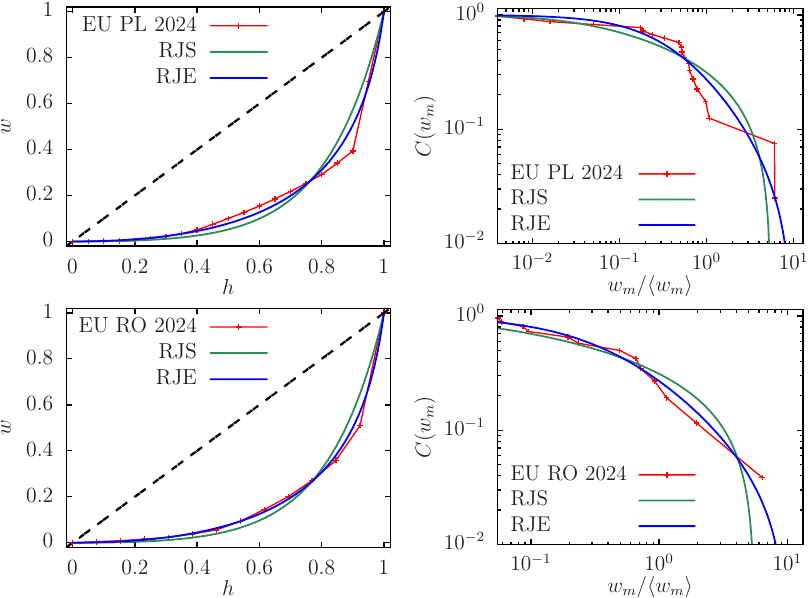}%
\end{center}
\caption{\label{figS5}
\label{fig_euvotePL2024}
As Fig.~2 for the EU election of PL/RO of 2024. 
For both cases also the theoretical RJE curves are shown (blue) and 
the optimal RJE parameters are $a=3.16,\ \eps_{\rm RJE}=0.102$ (EU PL 2024) 
and $a=3.22,\ \eps_{\rm RJE}=0.0982$ (EU RO 2024).
}
\end{figure}

\begin{figure}[h]
\begin{center}
\includegraphics[width=0.95\columnwidth]{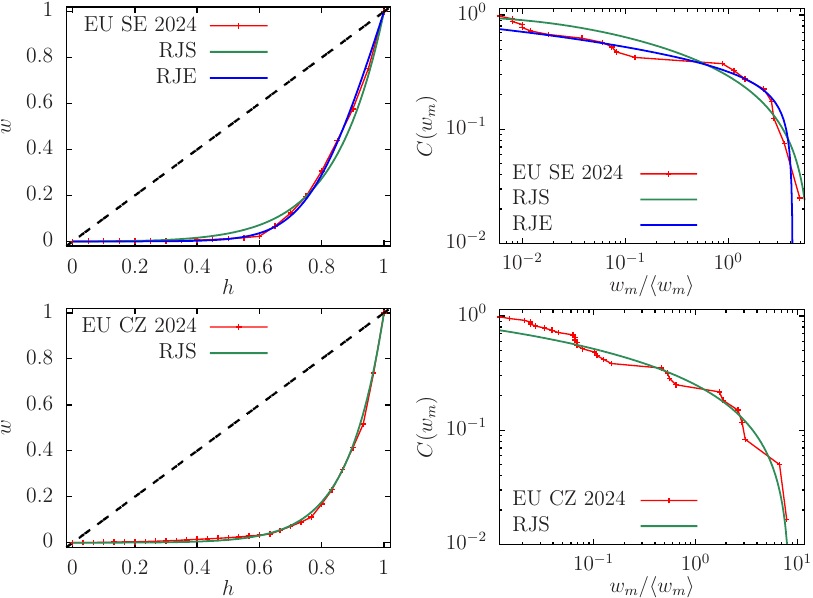}%
\end{center}
\caption{\label{figS6}
\label{fig_euvoteSE2024}
As Fig.~2 for the EU election of SE/CZ of 2024. 
For SE also the theoretical RJE curves are shown (blue) and 
the optimal RJE parameters are $a=-2.69,\ \eps_{\rm RJE}=0.236$.
}
\end{figure}

\begin{figure}[h]
\begin{center}
\includegraphics[width=0.95\columnwidth]{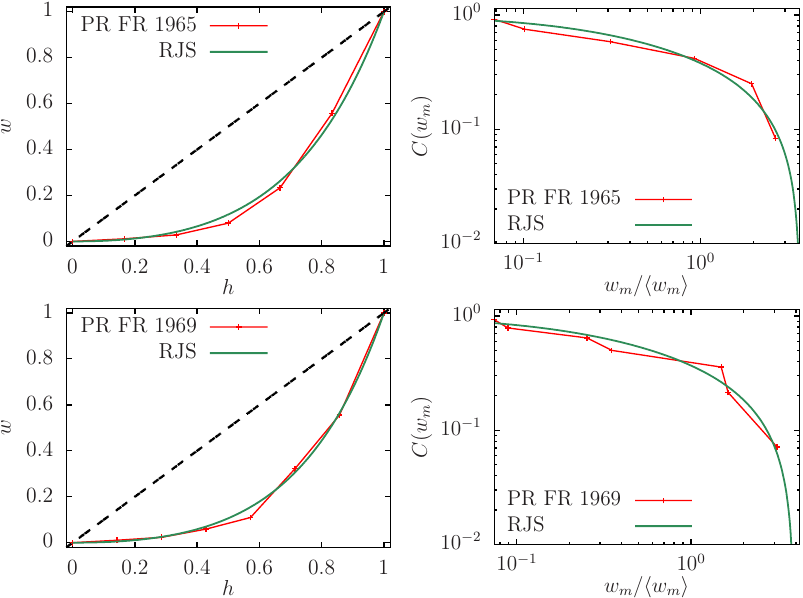}%
\end{center}
\caption{\label{figS7}
\label{fig_FR_pres_1965}
As Fig.~6 but for the first round FR presidential 
elections of 1965 and 1969. 
}
\end{figure}

\begin{figure}[h]
\begin{center}
\includegraphics[width=0.95\columnwidth]{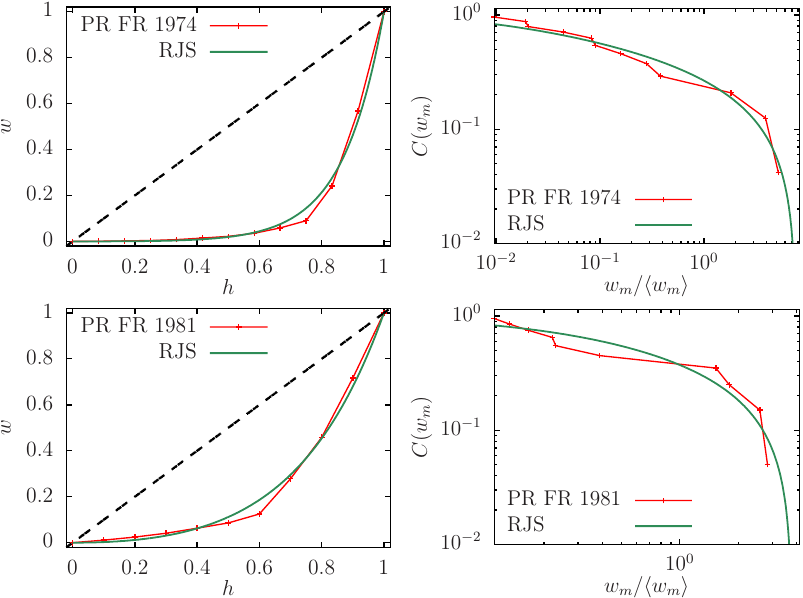}%
\end{center}
\caption{\label{figS8}
\label{fig_FR_pres_1974}
As Fig.~6 but for the first round FR presidential 
elections of 1974 and 1981. 
}
\end{figure}

\begin{figure}[h]
\begin{center}
\includegraphics[width=0.95\columnwidth]{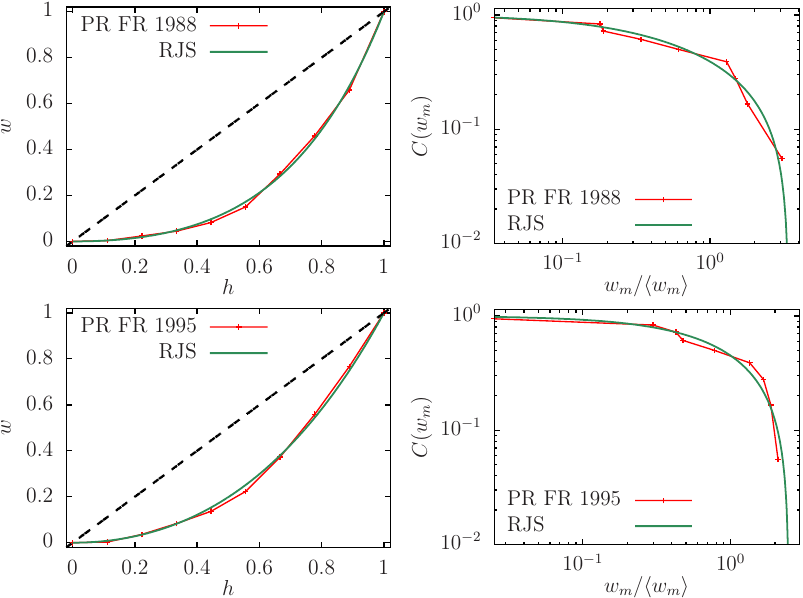}%
\end{center}
\caption{\label{figS9}
\label{fig_FR_pres_1988}
As Fig.~6 but for the first round FR presidential 
elections of 1988 and 1995.
}
\end{figure}

\begin{figure}[h]
\begin{center}
\includegraphics[width=0.95\columnwidth]{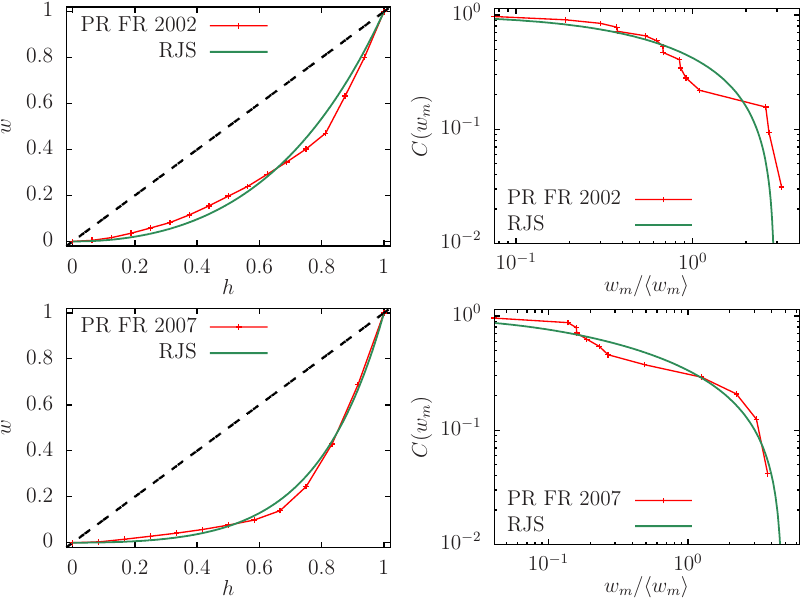}%
\end{center}
\caption{\label{figS10}
\label{fig_FR_pres_2002}
As Fig.~6 but for the first round FR presidential 
elections of 2002 and 2007.
}
\end{figure}

\begin{figure}[h]
\begin{center}
\includegraphics[width=0.95\columnwidth]{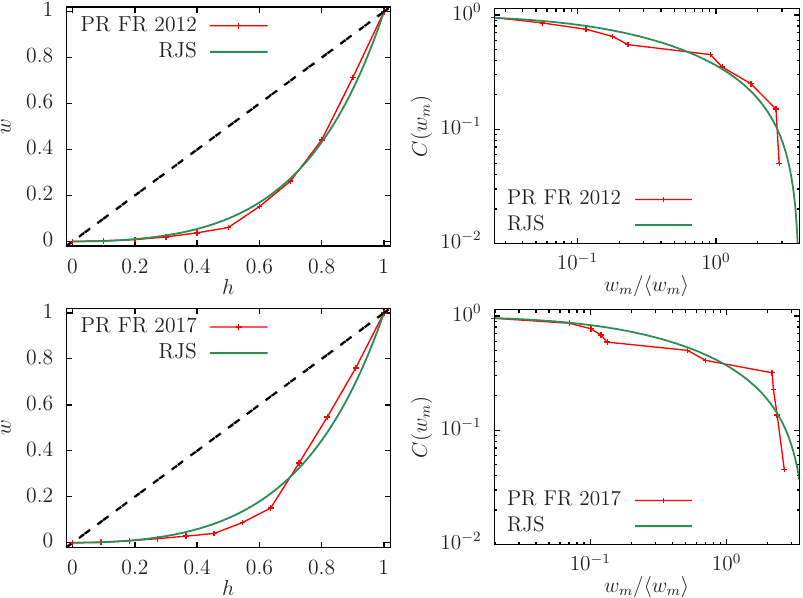}%
\end{center}
\caption{\label{figS11}
\label{fig_FR_pres_2012}
As Fig.~6 but for the first round FR presidential 
elections of 2012 and 2017.
}
\end{figure}

\begin{figure}[h]
\begin{center}
\includegraphics[width=0.95\columnwidth]{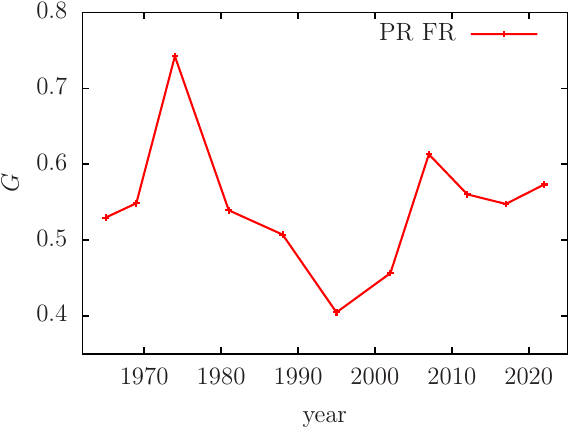}%
\end{center}
\caption{\label{figS12}
\label{fig_Gini_pres}
Dependence of the Gini coefficient 
of all FR first round presidential since 1965 on the election year 
(11 elections). 
}
\end{figure}

\end{document}